\documentstyle[aps,prl,epsf]{revtex}
\begin{document}
\title{Symmetric-Asymmetric transition in mixtures of Bose-Einstein condensates}
\author{Anatoly A. Svidzinsky and Siu-Tat Chui}
\address{Bartol Research Institute, University of Delaware, Newark, DE 19716}
\date{\today }
\maketitle

\begin{abstract}
We propose a new kind of quantum phase transition in phase separated
mixtures of Bose-Einstein condensates. In this transition, the distribution
of the two components changes from a symmetric to an asymmetric shape. We
discuss the nature of the phase transition, the role of interface tension
and the phase diagram. The symmetric to asymmetric transition is the
simplest quantum phase transition that one can imagine. Careful study of
this problem should provide us new insight into this burgeoning field of
discovery.
\end{abstract}


There is much recent interest in quantum phase transitions. Examples of
these include the Wigner electron solid melting transition, the Mott-Hubbard
metal-insulator transition, and different magnetic transitions. In physical
phenomena involving Bose-Einstein condensates (BECs) quantum mechanics play
a crucial role. In this paper we propose a new kind of quantum phase
transitions in phase-separated mixtures of BEC condensates. In this
transition, the distribution of the two components changes from a symmetric
to an asymmetric shape. To explore this transition, we first investigate the
stability of the symmetric phase by studying its normal modes. We find
interface modes that become soft. When the lowest frequency becomes zero,
the instability sets in and this determines the stability limit of the
symmetric phase. We determine the actual phase boundary by comparing the
energy between the symmetric phase and the asymmetric phase and find that
this actual phase boundary and the instability boundary is not the same.
This suggests that the transition is first order. The system may be a good
laboratory to study issues of quantum metastability and tunnelling. The
symmetric to asymmetric transition is the simplest quantum phase transition
that one can imagine. Careful study of this problem should provide us new
insight into this burgeoning field of discovery. We now describe the results
in detail.

Mixtures of trapped Bose-Einstein condensates have recently received
considerable theoretical~\cite{Ho96,Esry97,Pu98,Ohbe98,Chui99,chui3,Ao98}
and experimental\cite{Myat97,Hall98a,Sten98} interest. Experimentally at low
fields, the spin exchange process can occur in an optically trapped
condensate, leading to spin domains~\cite{Sten98} with metastable behavior~%
\cite{Mies99,Stam99b}. Binary condensates in two hyperfine levels of $^{87}$%
Rb have been created and studied~\cite{Myat97,Matt98}, most notably
realizing a system of interpenetrating Bose fluids~\cite{Hall98a},
measurements of phase dispersion~\cite{Hall98b}, and a vortex state in a
dilute-gas BEC~\cite{Matt99a}.

The equilibrium density distributions of segregated mixtures in the absence
of gravity have been studied numerically for different system parameters.
Two types of configurations have been discussed: a symmetric~\cite
{Pu98,Chui99,Esry97} configuration, for which one component is inside the
other one, and an asymmetric one in which the two components occupy the left
and the right hand side of a sphere~\cite{Chui99,Esry97}.

In general, the asymmetric phase possesses a lower interface energy. On the
other hand, since the degree of self-repulsion may differ between the two
species, the less self-repulsive component will prefer to remain where the
density is higher, while the other component moves to the low density
regions outside. This favors the symmetric phase. Depending on the system
parameters, one of these two energetic considerations will win out. These
system parameters can be adjusted by changing the trapping frequencies, the
relative particle numbers of the two species, and the interaction between
the particles with Feshbach resonances. We first address the stability of
the symmetric phase.

We consider the two-component BEC in a spherically symmetric trap. The
dynamics of this system is described by the time dependent Gross-Pitaevskii
equations:

\begin{equation}
\label{a1}i\hbar \frac{\partial \Psi _1}{\partial t}=-\frac{\hbar ^2}{2m}%
\Delta \Psi _1+V_{{\rm tr}}\Psi _1+\frac{4\pi \hbar ^2}m(a_{11}|\Psi
_1|^2+a_{12}|\Psi _2|^2)\Psi _1, 
\end{equation}
\begin{equation}
\label{a2}i\hbar \frac{\partial \Psi _2}{\partial t}=-\frac{\hbar ^2}{2m}%
\Delta \Psi _2+V_{{\rm tr}}\Psi _2+\frac{4\pi \hbar ^2}m(a_{22}|\Psi
_2|^2+a_{12}|\Psi _1|^2)\Psi _2, 
\end{equation}
where $\Psi _{1,2}$ are the condensate wave functions, $V_{{\rm tr}}=m\omega
_0^2r^2/2$ is the trapping potential, $\omega _0$ is the trapping frequency, 
$r$ is the radial spherical coordinate, $a_{ij}>0$ are $s-$wave scattering
lengths.

We shall assume in this paper that the condition $\label{a21}%
a_{12}^2-a_{11}a_{22}>0$ is satisfied and, therefore, the condensates are
phase-segregated. We study the condensates in the TF limit. In this regime,
the phase-segregated condensates overlap over the length scale $\Lambda =\xi
/\sqrt{a_{12}/\sqrt{a_{11}a_{22}}-1}$, where $\xi $ is the healing length 
\cite{Ao98}. For parameters of the JILA experiments on phase-segregated
states $\Lambda \approx 47\xi $. If the penetration depth $\Lambda \ll R$,
where $R$ is the size of the system, the condensates can be approximately
treated as nonoverlapping, which we assume to be the case. The effect of
overlapping condensates results in a finite surface tension and can be
included via boundary conditions at the interphase. If the condensates do
not overlap one can neglect the last terms in Eqs. (\ref{a1}), (\ref{a2}).
As a result, the dynamical equations for $\Psi _1$ and $\Psi _2$ decouple.
However, the two condensates are coupled by the boundary conditions at the
interface which require continuity of the pressure and the normal velocity.

The symmetric phase consists of a central core of the first component and an
outer shell of the second species (we assume $a_{11}\leq a_{22}$). The
stationary density distribution $n_i=|\Psi _i|^2$ of the two components is
given by 
\begin{equation}
\label{d1}G_{11}n_1=\mu _1(1-r^2/R_1^2),\quad 0<r<R_{*}, 
\end{equation}
\begin{equation}
\label{d2}G_{22}n_2=\mu _2(1-r^2/R_2^2),\quad R_{*}<r<R_2, 
\end{equation}
where $G_{ii}=4\pi \hbar ^2a_{ii}/m$, $R_i=\sqrt{2\mu _i/m\omega _0^2}$. The
normalization condition $\int n_idV=N_i$, where $N_i$ are the numbers of
condensate particles, determines the chemical potentials $\mu _i$. The
position of the phase boundary $R_{*}$ is given by the condition that the
pressures exerted by both condensates are equal $R_{*}=R_2\sqrt{(1-\kappa
\lambda )/(1-\kappa )\lambda }$ \cite{Chui02}, where we introduced the
dimensionless parameters $\kappa =\sqrt{a_{11}/a_{22}},$ $\lambda =\mu
_2/\mu _1$. The symmetric configuration is favorable when $\kappa $ differs
from unity, with the less repulsive component in the middle ( $\kappa <1$).
At the interface between two components $n_2/n_1=\kappa <1$.

One can rewrite the time dependent Gross-Pitaevskii equations in a
hydrodynamic form which shows an analogy between our problem and motion of
two immiscible fluids. In the strong phase-segregated regime the dynamics of
each components is described by the following Josephson hydrodynamic
equations \cite{Fett96}

\begin{equation}
\label{s1}\frac{\partial n}{\partial t}+\nabla \cdot (n{\bf V})=0, 
\end{equation}
\begin{equation}
\label{s2}\frac 12mV^2+V_{{\rm tr}}-\frac{\hbar ^2}{2m}\frac 1{\sqrt{n}}%
\Delta \sqrt{n}+Gn+m\frac{\partial \Phi }{\partial t}=\mu , 
\end{equation}
where ${\bf V}$ is the condensate velocity and $\Phi $ is the velocity
potential, ${\bf V=\nabla }\Phi $. The trapping potential plays the role of
gravitational potential in hydrodynamics. We look for small perturbation
from the equilibrium state. The equation for perturbation in the velocity
potential for each of the component is the same as the one component case 
\cite{Stri96} 
\begin{equation}
\label{e2}\frac{2\omega ^2}{\omega _0^2}\Phi -2r\frac{\partial \Phi }{%
\partial r}+(R^2-r^2)\Delta \Phi =0. 
\end{equation}
Continuity of the pressure at the interphase results in the following
boundary condition 
\begin{equation}
\label{s10}\omega _0^2R_{*}\frac{\partial \Phi _1}{\partial r}-\omega ^2\Phi
_1=\kappa \omega _0^2R_{*}\frac{\partial \Phi _2}{\partial r}-\kappa \omega
^2\Phi _2. 
\end{equation}
Continuity of $V_r$ gives another boundary condition at the interface 
\begin{equation}
\label{s11}\frac{\partial \Phi _1}{\partial r}=\frac{\partial \Phi _2}{%
\partial r}. 
\end{equation}
Eq. (\ref{e2}) and the boundary conditions (\ref{s10}), (\ref{s11})
constitute a complete set of equations necessary to determine the normal
modes of the system. We expand the velocity potential in terms of spherical
harmonics $\Phi =\Phi (r)Y_{lm}(\theta ,\phi )$. From (\ref{e2}) the radial
component can be written in terms of hypergeometric functions $\Phi
_1=C_1r^lF(\alpha ,\beta ,l+3/2,r^2/R_1^2)$, $\Phi _2=C_2r^lF(\alpha ,\beta
,1,1-r^2/R_2^2)$, where $\alpha +\beta =l+3/2,$ $\alpha \beta =(l-\omega
^2/\omega _0^2)/2$. From matching boundary conditions, we finally obtain the
eigenvalue equation for the normal mode frequencies 
\begin{equation}
\label{s16}\frac{\omega ^2}{\omega _0^2}=(1-\kappa )\frac{\left[
l(l+3/2)+(l-\omega ^2/\omega _0^2)\lambda xs_1(\omega ,x)\right] \left[
l-(l-\omega ^2/\omega _0^2)xs_2(\omega ,x)\right] }{\left[ l(l+3/2)(1-\kappa
)-x(l-\omega ^2/\omega _0^2)\left( \kappa \lambda s_1(\omega
,x)+(l+3/2)s_2(\omega ,x)\right) \right] }, 
\end{equation}
where $s_1(\omega ,x)=F(\alpha +1,\beta +1,l+5/2,\lambda x)/F(\alpha ,\beta
,l+3/2,\lambda x)$, $s_2(\omega ,x)=F(\alpha +1,\beta +1,2,1-x)/F(\alpha
,\beta ,1,1-x)$, $x=R_{*}^2/R_2^2$.

One special solutions of Eq. (\ref{s16}) is $\omega ^2=l\omega _0^2$, which
coincide with those for the one component condensate \cite{Stri96}. For this
solution the components oscillate in-phase and $\Phi _1=\kappa \Phi
_2\propto r^lY_{lm}(\theta ,\phi )$. Another special exact solution is $%
\omega ^2=5\omega _0^2$ with $\Phi _1\propto 1-5r^2/3R_1^2$, $\Phi _2\propto
\lambda (1-5r^2/3R_2^2)$. For this solution the components oscillate
out-of-phase if $\sqrt{3/5}R_1<R_{*}<\sqrt{3/5}R_2$ and in-phase otherwise.
In general we have solved Eq. (\ref{s16}) numerically and found the normal
mode frequencies $\omega $ of the two component condensate as a function of
the parameter $\kappa =\sqrt{a_{11}/a_{22}}$ for a fixed ratio $R_{*}/R_2$.
The ratio $R_{*}/R_2$ can be directly measured experimentally. Fig. 1 shows
the low frequency modes that become imaginary at $\kappa >1$. These modes
are peculiar for the two component systems and are analogous to the waves at
the interphase between two layers of immiscible fluids under gravity \cite
{Land88}. For this system, when the top layer becomes denser, the
gravitational energy becomes higher and the system becomes unstable. In a
similar manner as soon as $\kappa $ becomes greater than $1$, our system
becomes unstable.

In the region $|1-\kappa |\ll 1\,$the mode frequencies are small: $|\omega
|\ll \omega _0$. 
We obtain 
\begin{equation}
\label{s19}\omega ^2\approx \omega _0^2(1-\kappa )f(l,x), 
\end{equation}
where%
$$
f(l,x)=\frac{l\left( l+3/2+xs_1(0,x)\right) \left( xs_2(0,x)-1\right) }{%
x\left[ s_1(0,x)+(l+3/2)s_2(0,x)\right] }>0. 
$$
Eq. (\ref{s19}) describes the behavior of the low frequency modes in the
region close to the point of instability $\kappa =1$. In this region $\omega
\propto \sqrt{1-\kappa }$ and becomes imaginary when $\kappa >1$. We next
address the correction of this equation due to the finite overlap of the
wave function and the surface tension.

In the TF limit the interface tension results in small corrections of the
order of $\xi /R_2$ to the normal mode frequencies. However, for the low
frequency modes in the vicinity $\kappa \approx 1$ the effect of interface
tension is substantial because the mode frequencies themselves are close to
zero. The interface tension $\sigma $ modifies the boundary condition for
the pressure so that the pressure difference at the interface is equal to
the surface pressure $P_1-P_2=\sigma (1/r_1+1/r_2)$, where $r_1$, $r_2$ are
the principle radii of curvature. As a result, the ratio of densities at the
interphase is

$$
\frac{n_2}{n_1}=\sqrt{\kappa ^2-\frac{4\sigma }{G_{22}n_1^2R_{*}}}=\kappa _{%
{\rm eff}}. 
$$
The interface tension is given by \cite{Timm98} 
\begin{equation}
\label{s0}\sigma =\frac 4{\sqrt{3}}\sqrt{(\xi _1^2+\xi _2^2)\left[ a_{12}/%
\sqrt{a_{11}a_{22}}-1\right] }P, 
\end{equation}
where $\xi _i$ represents the single condensate coherence length $\xi
_i=\hbar /\sqrt{2m_iG_{ii}n_i}$, the pressure $P\approx G_{ii}n_i^2/2$ and
the condensate densities $n_i$ are estimated near the interface. Using the
modified boundary condition Eq. (\ref{s19}) is changed to 
\begin{equation}
\label{s23}\omega ^2\approx \omega _0^2\left( 1-\frac{3\sigma }{2R_{*}P}+%
\frac{\sigma (l-1)(l+2)}{mn_1\omega _0^2R_{*}^3}-\kappa _{{\rm eff}}\right)
f. 
\end{equation}
The interface tension shifts the frequencies of the lowest modes and changes
the stability region. For $l=0$ the inner droplet moves as a whole without
changing its shape. However, the displacement of the droplet into the less
dense region decreases the interface energy which is proportional to $%
n^{3/2} $. This is the origin for the contribution $-3\sigma/R_{*}P$ to the
mode frequency. As $\kappa $ increases the mode with $l=1$ becomes imaginary
first. This determines the system's stability limit. The two component
condensate is locally unstable when 
\begin{equation}
\label{s26}\frac{a_{11}}{a_{22}}>1-\frac \sigma {R_{*}P}=1-\frac{4\sqrt{6}%
\xi }{3R_{*}}(a_{12}/\sqrt{a_{11}a_{22}}-1)^{1/2}. 
\end{equation}

We next turn our attention to the calculation of the global stability
condition. In the asymmetric phase, we have component one on the right and
component two on the left (Fig. 2). In the TF approximation, the density
distribution is again given by Eqs. (\ref{d1}), (\ref{d2}). The position of
the interphase boundary between them is now different. We determine the
global stability condition by equating the energy of the symmetric and the
asymmetric phases. The calculation is similar to our previous work \cite
{Chui02}. In the limit $N_1\ll N_2$ we find a simple expression for the line
of global instability: 
\begin{equation}
\label{s27}\frac{a_{11}}{a_{22}}=1-4\left( \frac{\sqrt{6}\xi }{R_{*}}\right)
^{1/2}(a_{12}/\sqrt{a_{11}a_{22}}-1)^{1/4}\text{.} 
\end{equation}

In general, the calculation can only be carried out numerically. In Fig. 3
we plot the phase diagram that shows different stability regions of two
component condensates. In our estimates we take $\xi /R_{*}=0.01$. When $%
a_{12}<\sqrt{a_{11}a_{22}}$ the homogeneous binary mixture is a stable
state. Otherwise the two components are phase separated. In the phase
separated region for small ratios of $a_{11}/a_{22}$ the state $(1,2)$ with
the first component inside and the second outside is the only stable
configuration. As the ratio $a_{11}/a_{22}$ increases this configuration
becomes globally unstable when we cross the left (solid) curve. However, the
system is locally stable since the configuration corresponds to a local
minima of energy. As $a_{11}/a_{22}$ is further increased, we cross the line
of local stability (second solid line) and the $(1,2)$ state becomes
unstable towards transforming into a new stable asymmetric state. In between
the two solid lines, the system may tunnel quantum-mechanically from the $%
(1,2)$ state to the asymmetric phase. A possible scenario is via quantum
nucleation. How this takes place in detail is beyond the scope of the
present paper. Eventually the asymmetric phase becomes the $(2,1)$ state as $%
\kappa $ is further increased. If initially the system is prepared in the $%
(2,1)$ state then as the ratio $a_{11}/a_{22}$ is decreased this
configuration first becomes globally unstable when we cross the right dotted
line and locally unstable when $a_{11}/a_{22}$ is close enough to $1$.

In summary, we have elucidated in detail a new symmetric-asymmetric
transition in mixtures of BEC's. To simplify the numerical details, we have
presented our results assuming the trapping frequencies of the two
components are the same. Experimentally, the transition can be observed by
either changing the interaction strengths or the ratio of the trapping
frequencies.

This work was supported by NASA, Grant No. NAG8-1427.

\begin{figure}
\bigskip
\centerline{\epsfxsize=0.48\textwidth\epsfysize=0.48\textwidth
\epsfbox{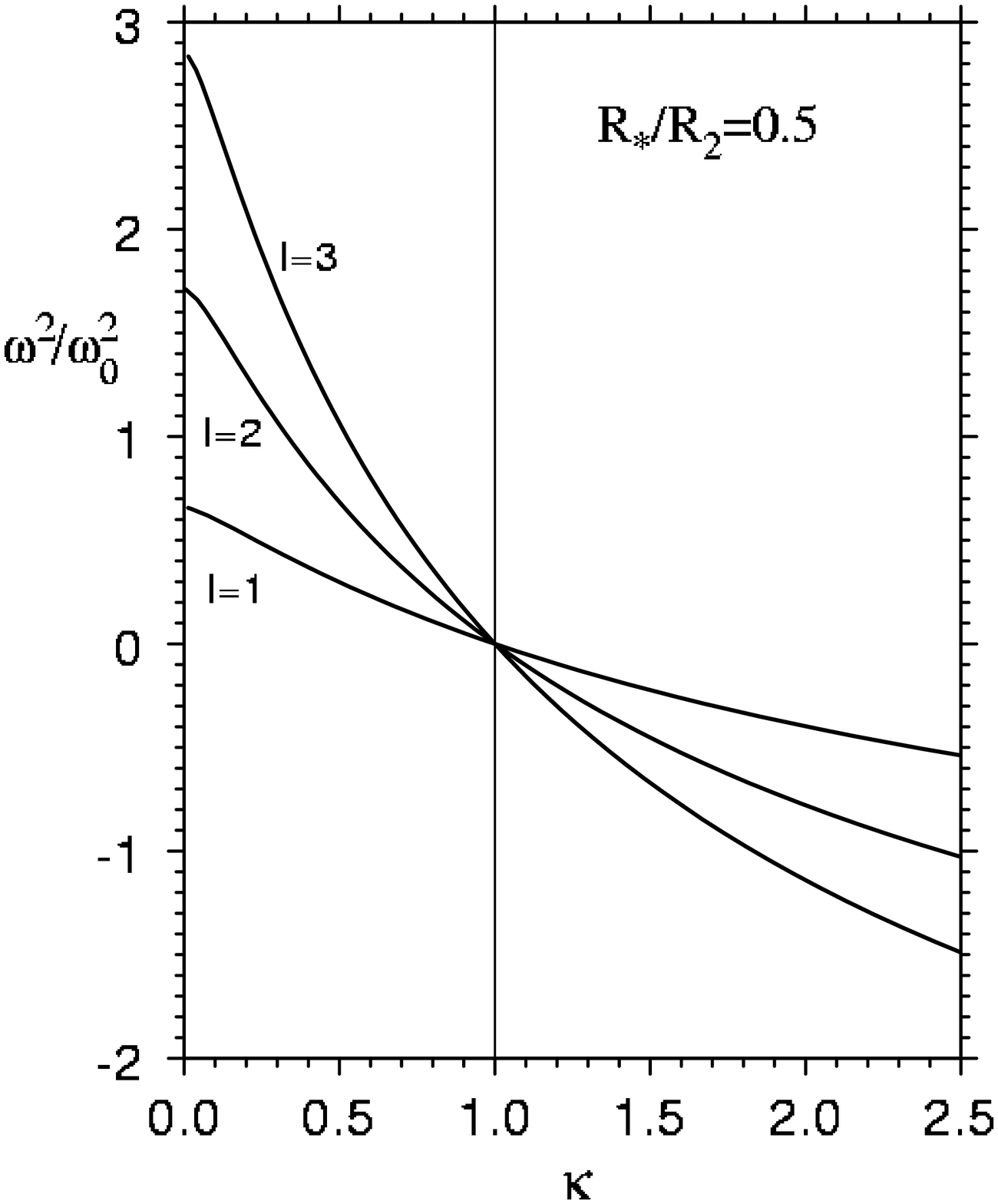}}

\label{fig1}
\end{figure}

{\small Fig. 1. Low frequency modes as a function of $\kappa =\sqrt{%
a_{11}/a_{22}}$ for different $l=1,2,3$. The position of the interface is $%
R_{*}=R_2/2$. Frequencies become imaginary at $\kappa >1$. }

\begin{figure}
\bigskip
\centerline{\epsfxsize=0.71\textwidth\epsfysize=0.4\textwidth
\epsfbox{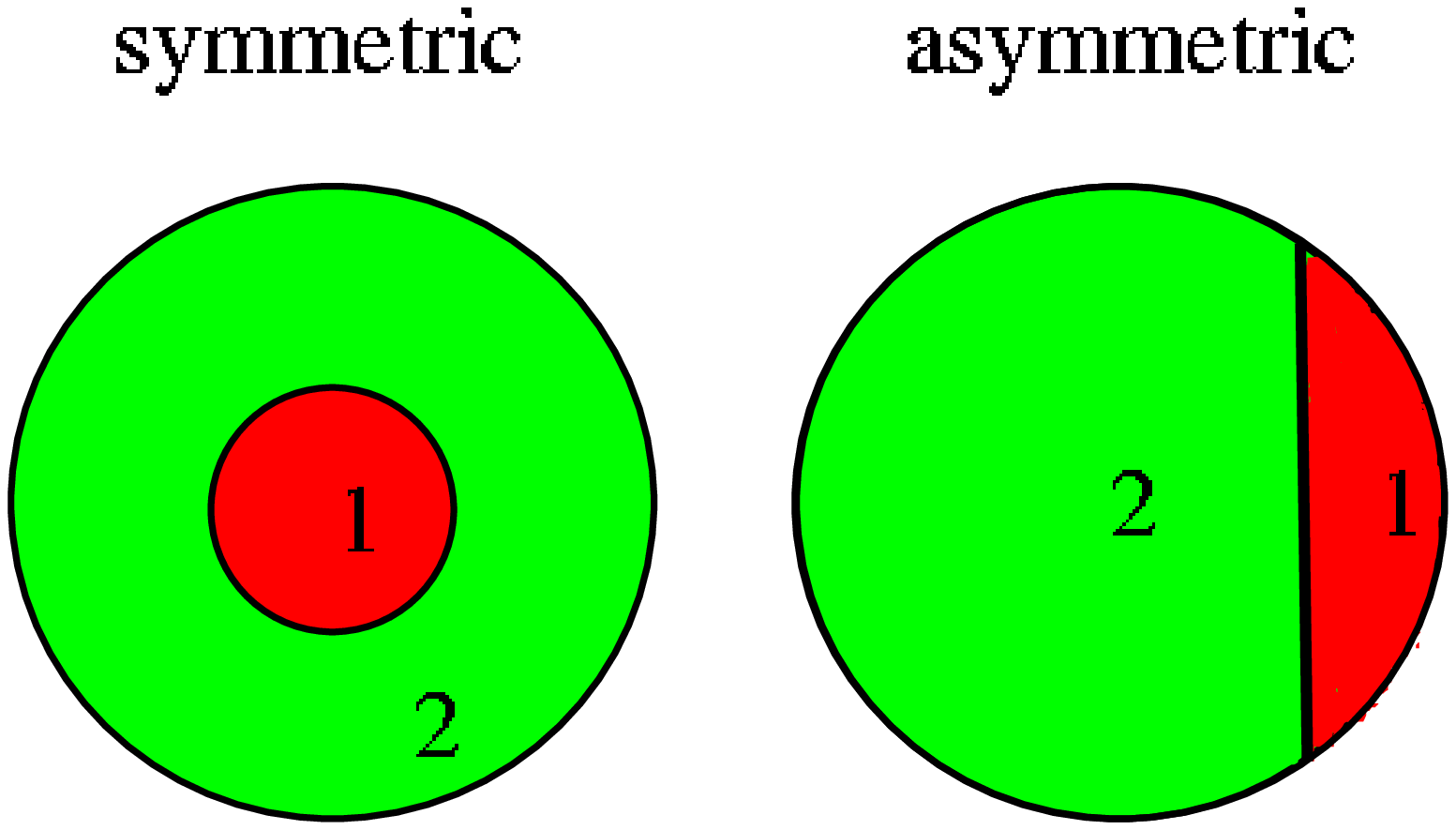}}

\label{fig2}
\end{figure}

\begin{center}
{\small Fig. 2. Symmetric and Asymmetric phases of two component BECs}
\end{center}

\begin{figure}
\bigskip
\centerline{\epsfxsize=0.48\textwidth\epsfysize=0.48\textwidth
\epsfbox{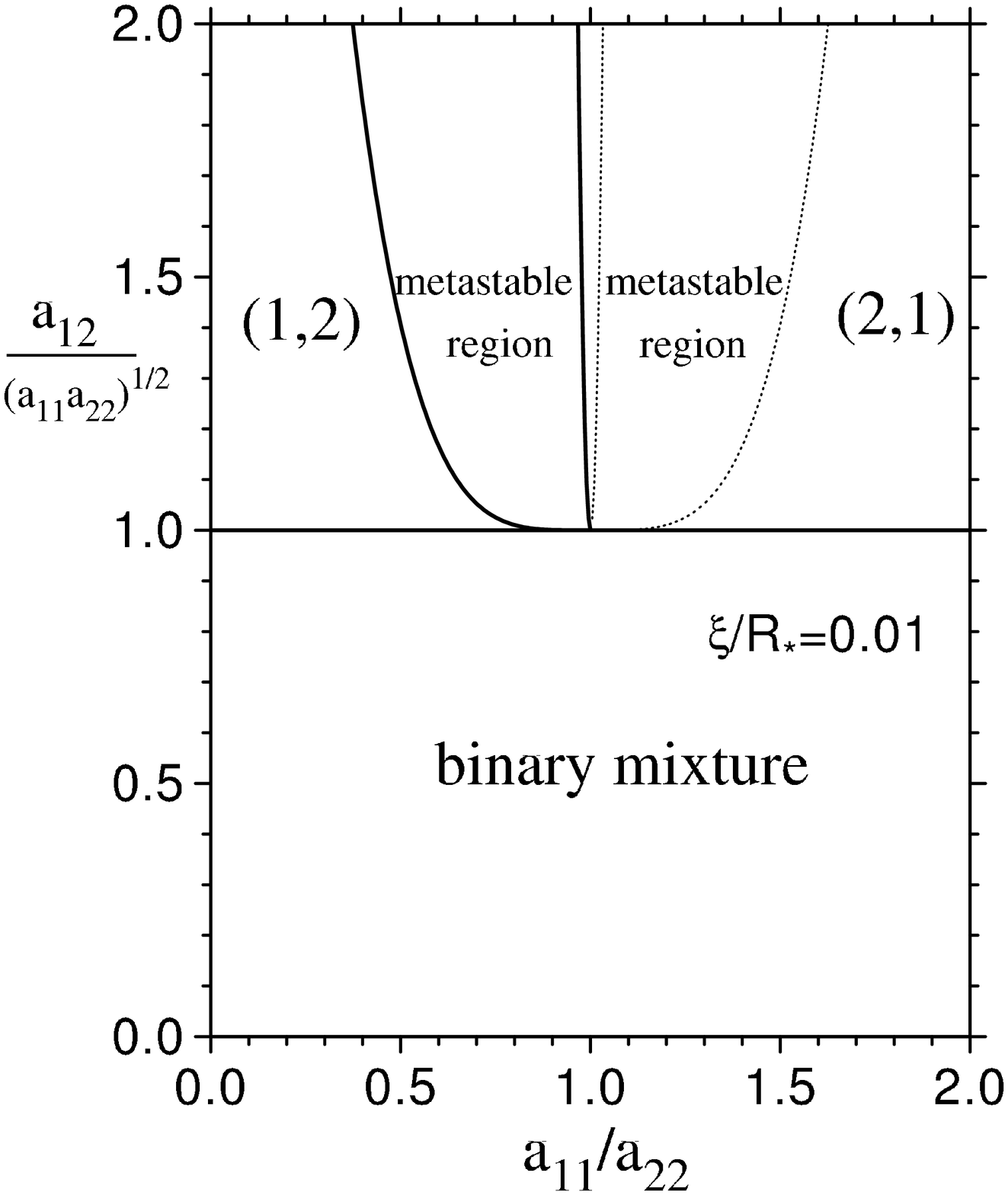}}

\label{fig3}
\end{figure}

{\small Fig. 3. Phase diagram of the two component condensate in coordinates
displaying relative interaction strength between bosons.}

\end{document}